\documentclass[preprint]{aastex}
\slugcomment{OU-TAP 161}
\shorttitle{Effects of Formation Epoch Distribution}
\shortauthors{Enoki et al.}

\begin{document}
\title{Effects of Formation Epoch Distribution on X-Ray Luminosity and 
Temperature Functions of Galaxy Clusters}

\author{Motohiro Enoki \altaffilmark{1,2}, Fumio Takahara \altaffilmark{1},
and Yutaka Fujita \altaffilmark{2}}

\altaffiltext{1}{Department of Earth and Space Science, Graduate School of Science, \\
Osaka University, Toyonaka, Osaka, 560-0043, Japan}

\altaffiltext{2}{National Astronomical Observatory, Osawa 2-21-1, Mitaka, Tokyo, 181-8588, Japan}


\begin{abstract}
We investigate statistical properties of galaxy clusters in the context
of hierarchical clustering scenario, taking account of their formation 
epoch distribution, motivated by the recent finding by Fujita
and Takahara that X-ray clusters form a fundamental plane, where the 
mass and formation epoch are regarded as two independent parameters.
Using the formalism which discriminates between major merger and 
accretion, the epoch of a cluster formation is identified with that of 
the last major merger. Since tiny mass accretion after the formation 
does not much affect the core structure of clusters, the properties 
of X-ray emission from clusters are determined by the total mass and 
density at their formation time. Under these assumptions, we calculate 
X-ray luminosity and temperature functions of galaxy clusters. We find 
that the behavior of luminosity function is different from the model 
which does not take account of formation epoch distribution, while 
the behavior of temperature function is not much changed. In our model, 
luminosity function is shifted to a higher luminosity and shows no 
significant evolution up to $z \sim 1$, independent of 
cosmological models. The clusters are populated on 
the temperature-luminosity plane with a finite dispersion. 
Since the simple scaling model in which the gas temperature is equal to 
the virial temperature fails to reproduce the observed 
luminosity-temperature relation, we also consider a model which takes 
the effects of preheating into account. The preheating model reproduces 
the observations much better.
\end{abstract}

\keywords{cosmology: theory --- clusters: galaxies: general --- X-rays: 
galaxies}

\section{Introduction}
\label{sec:intro}
Galaxy clusters are the largest virialized objects in the universe and 
should be useful cosmological probe since the properties of clusters 
are considered to depend on cosmological parameters and to reflect the 
structure formation history. Galaxy clusters contain a diffuse hot gas 
which emits X-rays via thermal bremsstrahlung. The X-ray temperature and
luminosity functions of clusters and their evolution have been used to 
determine the density of the universe and the amplitude of the rms 
density fluctuation. If there exist unique mass-temperature ($T$) and 
mass-luminosity ($L$) relations, one can predict X-ray temperature and
luminosity functions of clusters at redshift $z$ from a theoretical mass 
function. Therefore, it is important to investigate into correlations 
among physical quantities  of clusters and examine if such unique 
relations exist. Recently, \citet{FTa} have found that 
clusters at low redshifts ($z<0.1$) form a plane (the fundamental plane) 
in the three dimensional space ($\log \rho_{\rm gas,0}$, $\log r_c$, $\log T$),
where $\rho_{\rm gas,0}$ is the central gas density, and $r_c$ is the core radius 
of clusters. The data on the plane still have a correlation and form a 
band on the plane. The observed relation $L \propto T^{3}$ turns out 
to be the cross section of the band normal to the major axis. 
The existence of the fundamental plane implies that the
clusters form a two-parameter family, suggesting that the physical 
quantities of clusters are determined by the halo mass and density at 
the time of their formation and that no unique mass-temperature or 
mass-luminosity relation exists \citep{FTb}. In fact, using $N$-body 
simulations, 
\citet{NFW97} found that the structure of clusters is related to their
formation epoch, although they claimed that the clusters are a 
one-parameter family by assuming a unique relation between the mass 
and formation redshift. 
In this paper, motivated by the finding of the fundamental plane, 
we present a formulation of statistics of 
clusters in the context of hierarchical structure formation theory 
taking account of formation epoch distribution.

So far, \citet[hereafter PS]{PS74} mass function has been
widely used to calculate temperature and luminosity functions 
(e.g. \citealp{Eke}). Although PS mass function provides the 
number of dark halos at a given time, it does not contain information about 
the formation time of halos. Thus, a conventional way to compute the 
temperature and luminosity functions is simply to assume that 
the formation time is the same as the observed time. Since in this case 
clusters form a one-parameter family about the mass, an extension of PS 
theory is necessary to investigate the two-parameter nature of clusters, 
taking account of the effect of the formation epoch distribution.
Using the merger probabilities in an extended Press \& Schechter
clustering model \citep{BCEK,Bow}, \citet[hereafter LC]{LC} derived a 
formation epoch distribution function in an enlightening way,  
although LC did not calculate predictions of X-ray luminosity and 
temperature functions.
The formation time in their model is defined by that time when the halo 
mass becomes half that at the observed epoch
and this definition does not discriminate between tiny and notable relative mass
capture, or between accretion and merger.
In the hierarchical clustering scenario, low-mass objects successively 
merge with one another to build up ever more massive objects. However, 
major deviations from equilibrium and subsequent relaxation take place 
only when halos of comparable masses merge, while tiny mass
captures have little effect on the capturing halos. 
So, it is desirable to devise such formulations. 
\citet{KSa,KSb} have attempted to describe the formation and 
destruction of halos within the extended PS prescription by 
discriminating between accretion and merger.
They define the halo formation and destruction rates and define the 
formation epoch distribution by utilizing the survival probability. 
Although they predict various statistical properties
of X-ray clusters, they assume an empirical relation between
temperature and luminosity in such predictions, which is not 
satisfactory from our point of view.

In this paper, we adopt another model of the formation epoch
distribution proposed by \citet[hereafter SSM]{SSM}. They developed 
a modification of the extended PS model that differentiates merger 
from accretion and define the formation epoch as the epoch
of the last major merger by utilizing the empirical mass-density
correlation obtained by $N$-body simulations. 
Following this formalism, we can obtain not only formation 
time distribution but also the halo mass at the formation time. 
In order to obtain statistics of galaxy clusters, we assume that
the halo mass and density at its last major merger determine the
structure of clusters. The temperature and luminosity of 
galaxy clusters are calculated in terms of the halo mass and density at 
the formation without resorting to empirical relations.
On the basis of these prescriptions, we construct a simple scaling model of 
X-ray clusters and calculate $L-T$ distribution, 
the temperature and luminosity functions.
Since it is well known that a simple scaling model between gas and dark
matter results in an $L-T$ relation different from observations 
\citep{Eke98}, we also investigate a simple preheating model.
The paper is organized as follows. The formulation of statistics is
presented in {\S}
\ref{Formulation}. The X-ray cluster model is constructed in {\S} 
\ref{X-ray}. The results and discussion are described in {\S} 
\ref{Results}.

\section{Formulation}\label{Formulation}
\subsection{Formation time distribution}
\label{Formation time distribution}

In this subsection, we summarize the SSM formalism. To follow the
formation and evolution of halos, SSM used a modified version of
the extended PS clustering model \citep[; LC]{BCEK,Bow}
and made a schematic distinction between minor and
major mergers by defining the formation of a halo as the last major
merger it experienced.
This definition does not affect the abundance of halos at a given time,
although it affects the description of their growth. Thus, the mass
function is equal to the PS mass function
\begin{equation}
n(M,t)dM = \sqrt{\frac{2}{\pi}} \frac{\rho_0}{M} 
\frac{\delta_c(t)}{\sigma^{2}(M)}
\left|  \frac{d\sigma(M)}{dM} \right| 
\exp \left[-\frac{1}{2} \frac{\delta_c^{2}(t)}{\sigma^{2}(M)} \right] dM,
\label{eq:PS}
\end{equation}
where $\rho_0$ is the present mean density of the universe,
$\delta_c(t)$ is the critical density contrast for collapse at $t$, and
$\sigma(M)$ is the rms density fluctuation in spheres containing a mean
mass $M$. In this paper, we use an approximate formula of
$\delta_c(t)$ for spatially flat cosmological model \citep{NS}
and a fitting formula of $\sigma(M)$ for the CDM fluctuation
spectrum \citep{Kphd}.

In the LC model, the instantaneous merger rate for halos with mass $M$ at
$t$ per infinitesimal range of final mass $M' > M$, or
specific merger rate, is
\begin{eqnarray}
r_{\rm LC}^{m}(M \to M',t) dM' & \equiv & \sqrt{\frac{2}{\pi}}
\ \left|  \frac{d\delta_c(t)}{dt} \right|  \ \frac{1}{\sigma^{2}(M')}
\ \left|  \frac{d\sigma(M')}{dM'} \right|  \nonumber \\
& \times & \left[1- \frac{\sigma^{2}(M')}{\sigma^{2}(M)} \right]^{-
3/2} \nonumber \\
& \times & \exp \left\{ -\frac{\delta_c^{2}(t)}{2} \left
[\frac{1}{\sigma^{2}(M')}-\frac{1}{\sigma^{2}(M)} \right] \right\} 
dM' .
\label{eq:LCmerge}
\end{eqnarray}

In the SSM formalism, it is assumed that a halo with mass $M$ experiences a major 
merger and is destroyed when the relative mass increment 
$\Delta M /M \equiv (M'-M)/M$ exceeds a certain threshold $\Delta_m$. 
The major merger is regarded as the formation of a new halo. 
On the other hand, when $\Delta M /M <\Delta_m$, the event is regarded 
as continuous accretion;
the halo keeps its identity and its core structure. Thus, from the
specific merger rate (eq. [\ref{eq:LCmerge}]), the mass accretion
rate, $R_{\rm mass}(M,t) \equiv dM/dt$, of halos with mass $M$ at time
$t$ is defined as
\begin{equation}
R_{\rm mass}(M,t) 
= \int_{M}^{M(1+\Delta_m)} \Delta M r_{\rm LC}^{m}(M \to M \acute{
} ,t) dM' .
\label{eq:R-mass}
\end{equation}
The destruction rate is defined as
\begin{equation}
r^{d}(M,t) = \int_{M(1+\Delta_m)}^{\infty} r_{\rm LC}^{m}(M \to M',t) dM
\acute{} .
\label{eq:destrate}
\end{equation}
The formation rate should be written by
\begin{equation}
r^{f}[M(t),t] = \frac{d \ln n[M(t),t]}{d t} + r^{d}[M(t),t] 
+ \partial_M R_{\rm mass}[M(t),t]
\label{eq:continue}
\end{equation}
from the conservation equation for the number density of halos per unit
mass along mean mass accretion tracks, $M(t)$, which is the solution of
the differential equation
\begin{equation}
\frac{dM}{dt} = R_{\rm mass}[M(t),t] .
\label{eq:M-track}
\end{equation}

From this formation rate, one can obtain the distribution of formation
time, $t_f$, for halos with masses $M$ at $t$;
\begin{equation}
\Phi_f(t_f;M,t)dt_f
= r^{f}[M(t_f),t_f] \exp \left\{- \int_{t_f}^{t} r_f[M(t'),t'] d
t\acute{} \right\} dt_f.
\label{eq:distSSM}
\end{equation}
From equation (\ref{eq:M-track}), the halo mass at formation
time, $M_f = M(t_f)$, becomes
\begin{equation}
M_f = M - \int_{t_f}^{t} R_{\rm mass}[M(t'),t'] dt'.
\label{eq:m_f}
\end{equation}
The value of $\Delta_m$ is fixed by the fits to the empirical
mass-density (or mass-radius) correlation obtained by $N$-body
simulations \citep{NFW96,NFW97}.  The best fit is $\Delta_m = 0.6$ in a
number of different cosmological models (SSM). To be precise, $M_f$ for
fixed $M$ and $z_f$ should have scatter around the mean accretion track
(eq. [\ref{eq:M-track}]). Therefore, $M_f$ in equation (\ref{eq:m_f}) is
an approximation to the mean formation mass of true
distribution. Although we ignore this scatter for the sake of
simplicity, it could be estimated by using the algorithm given in
\citet{NuSh}.

In this paper,
we investigate two cosmological models (SCDM and LCDM). In Table
\ref{parameter}, we tabulate the cosmological density parameter
($\Omega_0$), the cosmological constant ($\lambda_0$), the Hubble
constant in units of $100 \rm\; km\; s^{-1}\;Mpc^{-1}$ ($h$), and
the present rms density fluctuation in spheres of radius $8 h^{-1}
\mbox{Mpc}$ ($\sigma_8$). Figure \ref{dist} shows the formation time
distributions (eq. [\ref{eq:distSSM}]) for the SCDM and LCDM models
for several values of the present halo mass.  Since in the LCDM model the
universe is in the accelerated expanding stage at $z \sim 0$, the growth
of density fluctuation has stopped. Thus, the slope of the formation
time distribution in the LCDM model is less steep in comparison with the
SCDM model at $z\sim 0$ (Figure \ref{dist}) . Figure \ref{mass} shows
halo masses at its formation time (eq. [\ref{eq:m_f}]) for the halo
with the present mass of $10^{15} M_{\odot}$.

Note that, in the PS formalism, there is no distinction between merger
and accretion.  Any mass capture is regarded as merger, that is, $t_f =
t$.  Consequently, the formation time distribution is
\begin{equation}
\Phi_f(t_f;M,t) = \delta^{D} (t_f - t)\:,
\label{eq:distPS}
\end{equation}
where $\delta^{D}(x)$ is Dirac's delta function.

\subsection{X-ray cluster model}\label{X-ray}

In the SSM formalism, the mass function of dark halos at a fixed epoch 
is given by the PS mass function. On the other hand, for a given mass,
the dark halos take a range of formation time as depicted by equation
(\ref{eq:distSSM}). Thus, dark halos form a two-parameter family. SSM
found that the characteristic density and the scale radius of a halo at
the present epoch are proportional to the critical density of the
universe and the virial radius of the halo at the formation time 
$t=t_f, $ respectively. This means that, between major mergers, 
halos gradually grow through the accretion of surrounding matter 
while keeping the central part unchanged. Considering the 
two-parameter family nature of clusters, we thus assume that the 
physical quantities of clusters of galaxies are represented by the 
mass and the virial density at the their formation time or the last  
major merger. On this assumption, the temperature and luminosity of a 
cluster are determined
by the formation redshift $z_f$ and the halo mass $M_f$ at $z_f$;
$T=T[z_f,M_f(M,z_f)]$, $L=L[z_f,M_f( M,z_f)]$.  Here, the relation
between $t_f$ and $z_f$ are straightforwardly determined once the
cosmological model is specified. In order to obtain $L$ and $T$, we use
the spherical collapse model \citep{T69,GG72} and assume that the
cluster is spherically symmetric and intracluster gas is in an 
isothermal hydrostatic equilibrium in the gravitational potential 
of dark matter halo.

In the spherical collapse model, the virial density $\rho_f$ of a halo
at the formation redshift $z_f$ is given by
\begin{equation}
\rho_f = \rho_c(z_{f}) \Delta_c(z_{f}) 
= \rho_{c0} \Delta_c(z_{f}) \frac{\Omega_0 (1+z_f)^
3}{\Omega(z_{f})},
\label{eq:density_c}
\end{equation}
where $\rho_c(z_{f})$ is the critical density of the universe,
$\Delta_c(z_{f})$ is the ratio of the virial density to the critical
density, and $\Omega(z_{f})$ is the cosmological density parameter. The
index 0 refers to the values at $z=0$. We use the fitting
formula of \citet{BN} for the virial density of spatially
flat cosmological models;
\begin{eqnarray}
\Delta_c &=& 18 \pi^{2} +82x-39x^{2}\:,
\end{eqnarray}
where $x \equiv \Omega(z_{f}) - 1 = (\Omega_0-1)/ 
[\Omega_0(1+z_f)^3+1-\Omega_0 ]$.  

The virial radius is obtained by
\begin{equation}
r_{f} = \left(\frac{3 M_{f}}{4 \pi \rho_{f}} \right)^{1/3}.
\label{eq:radius}
\end{equation}
The virial temperature $T_{\rm vir}$ is given by
\begin{eqnarray}
k_B T_{\rm vir} &=& \frac{\mu m_p GM_{f}}{3 r_{f}} \nonumber \\
&=& 0.92 \ \left(\frac{M_{f}}{10^{15} M_{\odot}} \right)^{2/3}
\left[h^2 \Delta_c \frac{\Omega_0}{\Omega(z_f)} \right]^{1/3} (1+z_f)
\ \ \mbox{keV}
\label{eq:Temp}
\end{eqnarray}
where $\mu$ is the mean molecular weight which we take to be $\mu=0.6$,
$m_p$ is the proton mass, and $k_B$ is the Boltzmann constant.

The luminosity of a cluster is
\begin{equation}
L = 6.50 \times 10^{-24} 
\left( \frac{k_B T}{1 \mbox{keV}} \right)^{1/2} \  
\int_0^{r_f} n_e^{2} 4 \pi r^{2} dr \ \mbox{erg s}^{-1}
\label{eq:X-lum}
\end{equation}
where $n_e$ is the electron number density in CGS units.  Note that 
the gas temperature $T$ is not equal to the virial temperature 
$T_{\rm vir}$ in general. In order to calculate luminosity from equation
(\ref{eq:X-lum}), we need to specify the gas density profile. Here, we
adopt isothermal $\beta$ model 
\begin{equation}
\rho_{\rm gas}(r) = \rho_{\rm gas,0} 
\left[1 + \left(\frac{r}{r_c} \right)^{2} \right]^{-3\beta/2}, 
\label{eq:Bfit}
\end{equation}
where $\rho_{\rm gas,0}$ is the central gas density and $r_c$ is core
radius.  The central gas density is calculated through the relation
\begin{equation}
\int_{0}^{r_f} \rho_{\rm gas} (r) 4 \pi r^{2} dr = M_{\rm gas}
\label{eq:gasfrac}
\end{equation}
where $M_{\rm gas}$ is the total gas mass. From now on, we assume that
$r_c = r_f/8 $ \citep{FTb}.

Thus, from (\ref{eq:X-lum}) and (\ref{eq:Bfit}), the luminosity is
given by
\begin{equation}
L = 1.79 \times 10^{44} 
\left(\frac{k_B T}{1 \mbox{keV}} \right)^{1/2} \left(\frac{M_f}{10^{1
5} M_{\odot}} \right)
\left[ h^{2} \Delta_c \frac{\Omega_0}
{\Omega(z_f)} \right] (1+z_f)^{3} f_
{m}^{2}
\ \mbox{erg s}^{-1}
\label{eq:lumino}
\end{equation}
where $f_m \equiv M_{\rm gas}/M_f$ is the gas mass fraction and we
assume here $\beta = 2/3$ for simplicity.  Equations (\ref{eq:Temp})
and (\ref{eq:lumino}) show that the temperature and luminosity are
functions of the halo mass and the formation redshift indeed when one
specifies the cosmological model and the relation between $T$ and
$T_{\rm vir}$.

\subsection{Temperature and luminosity function}\label{TL}

Next, we formulate statistics of galaxy clusters.  
Now that the temperature and luminosity are expressed as 
functions of $t_f$ and $M_f$, such that $T=T[t_f,M_f(M,t_f)]$ and
$L=L[t_f,M_f(M,t_f)]$, we can derive temperature and luminosity
functions using the transformation of the two variables from $(t_f,M)$
to $(T,L)$ as follows.

For a mass range $M \sim M +dM$ at $t$, the comoving number density of
clusters formed at $t_f \sim t_f + dt_f$ is given by
\begin{equation}
n(t_f,M;t) dt_f dM = n(M;t) \Phi_f(t_f;M,t)dt_fdM.
\end{equation}
Thus the comoving number density of clusters at $t$ with $T \sim T+dT$
and $L \sim L+dL$ is given by
\begin{eqnarray}
n(T,L;t)dTdL &=& n(t_f,M;t)dt_fdM .
\end{eqnarray}
Accordingly,
\begin{eqnarray}
n(T,L;t) &=& n(t_f,M;t) \left| \frac{\partial(t_f,M)}{\partial(T,L)} \right|.
\end{eqnarray}
Since the comoving number density $n(T,L;t)$ gives the distribution 
function on $L-T$ plane, it reflects $L-T$ relation of clusters.  

The temperature function and luminosity function are
respectively given by
\begin{eqnarray}
n(T;t) &=& \int n(T,L;t) dL
\end{eqnarray}
and
\begin{eqnarray}
n(L;t) &=& \int n(T,L;t) dT.
\end{eqnarray}
It is to be noted that these functions take account of the distribution
of cluster formation time, $t_f$.

Thus, given the relations $T = T[t_f,M_f(M,t_f)]$ and 
$L = L[t_f,M_f(M,t_f)]$,
one can calculate the $L-T$ distribution function, the temperature
function and the luminosity function.

\section{Results and discussion}\label{Results}

In this section, we derive the temperature function, the luminosity
function and $L-T$ distribution function. We assume that the gas mass
fraction of clusters $f_m$ is the same as the cosmic baryon ratio
$\Omega_b/\Omega_0$ where $\Omega_b$ is the density parameter of baryon
and we adopt $\Omega_b = 0.0125h^{-2}$ to be consistent with the
primordial nucleosynthesis.

\subsection{Scaling model}\label{R-1}

First, we examine the case where the gas temperature $T$ is equal to the
virial temperature $T_{\rm vir}$. In this case, the structure is
self-similar and we call this model the scaling model. We calculate the
temperature and luminosity functions based on the SSM formalism
(eq. [\ref{eq:distSSM}]) and compare them with those based on the PS
formalism (eq. [\ref{eq:distPS}]).  The temperature function and their
redshift evolution are shown in Figure~\ref{nT}, where thin lines and
symbols denote SSM and PS predictions, respectively.  As is seen, there
is little difference between the predictions of the two formalisms.
This can be explained as follows. When we take account of the
distribution of the formation time, physical quantities of clusters are
affected by two factors.  First, if the object formed earlier, its
virial density becomes larger (eq. [\ref{eq:density_c}]).  Second, if
the object formed earlier, its mass at the formation becomes smaller for
a given mass at $z=0$ (Figure~\ref{mass}).  Gas temperature depends on
the virial density and mass at the formation time as $T \propto M_f/r_f
\propto \rho_f^{1/3} M_f^{2/3}$, and these two factors tend to cancel
out. This behavior of temperature function is the same as that of
previous studies (e.g. \citealp{KSb}).  This feature is common to both
the cosmological models we investigate. The $z_f$ dependence of
temperature, $T=T(z_f,M = 10^{15} M_{\odot})$, is shown in Figure
\ref{T-z}, which explicitly shows that $T$ is roughly constant for $z_f
\la 2$.  Note that \citet{Ma} also finds no evidence for a correlation
between X-ray temperature and formation time in his simulation of X-ray
clusters. The difference between SCDM model and LCDM model is mainly
caused by the dependence on the time evolution of virial density
(eq. [\ref{eq:density_c}]).

The luminosity function and their redshift evolution are shown in
Figure \ref{nL}, where thin lines and symbols denote SSM and 
PS predictions, respectively. As is seen, SSM formalism predicts larger 
number density than PS formalism and for the SSM formalism, there is 
little evolution even in SCDM model in contrast with the PS formalism.  
This result is different from that of the temperature function, because of
the difference between the dependence of temperature and luminosity on
the virial density and mass at the formation. Since $L \propto
\rho_f^{2} r_f^{3} T^{1/2} \propto \rho_f M_f T^{1/2} \propto
\rho_f^{7/6} M_f^{4/3}$, the luminosity depends more strongly on the
virial density than the temperature. The $z_f$ dependence of luminosity,
$L=L(z_f,M = 10^{15} M_{\odot})$, is presented in Figure \ref{L-z},
which shows that $L$ increases with $z_f$.  
Therefore, earlier formed dense
clusters contribute to the increase of the luminosity function shown
in Figure \ref{nL}.  Moreover, the increase of $L$ with $z_f$ explains 
that there is little evolution of the luminosity function from $z=1.0$ 
to $z=0$ both in SCDM and LCDM universes. 
Since the growth of density fluctuations in SCDM model is more rapid 
than that in LCDM model, one might think
that the luminosity function in SCDM model should evolve rapidly. 
However, the present result implies that it is not the case if we 
consider the effect of the distribution in the formation redshift.  
Thus, recent observational evidence for little evolution of luminosity
function \citep{O1,O2,O3} does not
necessarily mean that SCDM model is disfavored against LCDM model.
It is also noted that if we consider the effect of formation epoch 
distribution, smaller value of $\sigma_8$ is needed 
to reproduce observations compared with the PS formalism.

When we compare the predictions with observations, 
one should note that the amplitude of temperature and luminosity 
functions can be adjusted by changing $\sigma_8$ so that 
we are concerned with their shape. 
In Figure \ref{nT}, the thick line is the best power-law fit to the
observed low redshift temperature function obtained by \citet{OnT}. 
The predicted shape of the temperature function is consistent with 
observations both for SCDM and LCDM models, although agreement is 
better for LCDM model.
In Figure \ref{nL} the thick line is the best fitted  
Schechter function to the observed bolometric luminosity function 
within $z=0.3$ obtained by \citet{OnL}. 
The predicted shape is much steeper than the observations both for 
SCDM and LCDM models. This is another representation of the well-known 
discrepancy of the $L-T$ relation. 

Next, we investigate the distribution on the $L-T$ relation.  For
several values of $n[\log(T),\log(L);z=0]$, we plot iso-density contours
on the $L-T$ plane in Figure \ref{L-T}. From the relations $T \propto
\rho_f^{1/3} M_f^{2/3}$ and $L\propto \rho_f^{7/6} M_f^{4/3}$, the
luminosity behaves as $L \propto \rho_f^{1/2} T^{2}$.  As discussed in
\S\ref{sec:intro}, in the PS formalism it is assumed that the observed
redshift of a cluster is equal to the formation redshift
($z=z_f$). Thus, most of the observed clusters ($z\sim 0$) have nearly
the same virial density $\rho_f (z)$ and their physical quantities
depend only on mass.  Therefore, in the PS formalism, clusters form a
one-parameter family, which is shown by the straight line.  On the other
hand, in the SSM formalism, the virial densities take a wide range of
values because $\rho_f$ depends on the formation redshift.  Thus,
clusters form a two-parameter family.  The scatter of the $L-T$ relation
shown in Figure~\ref{L-T} reflects the dispersion of the halo formation
time.  \citet{SM} also pointed this out.  Since $\rho_f$ tends to be
distributed more widely for smaller clusters, the scatter of $L-T$
relation is larger at lower $T$ and $L$.  Thus, the slope of $L-T$
relation in the SSM formalism becomes shallower than $L \propto T^{2}$.
This is, however, in conflict with the observed correlation $L \propto
T^{3}$ (e.g. \citealp{oLT}).  As long as we assume that the ratio of the
gas density of the core to the virial density is constant, this tendency
persists.  To resolve this discrepancy, this ratio should vary such that
less massive clusters have smaller baryon fraction at the cluster core
from which much of the X-ray emission originates
\citep{met94,kay99,val99,wu99}. Many authors have attributed such
behavior to preheating of intracluster gas
\citep{kai91,evr91,CMT,bal99}, in which intracluster gas had already
been heated before the cluster formed. On the other hand, some authors
claimed that intracluster gas is heated after the formation of the
cluster \citep{loe00,bri01}.  Moreover, \citet{Br} argued that cooling
of gas and the resultant galaxy formation are sufficient to explain to
lower the gas fraction in small clusters and groups without additional
heating.  Since these models give qualitatively similar gas
distributions, we will adopt a preheating model in the next subsection.
It is to be noted future observations may discriminate the heating or
galaxy formation models \citep[e.g.][]{fuj01}.

\subsection{Preheating model}\label{R-2}

We examine the effects of preheating using a simple model of
\citet{FT00} based on the models of \citet{CMT}.
They combine effects of shock heating and preheating so that
the gas temperature is higher than the virial temperature such that 
\begin{equation}
T = T_{\rm vir} + \frac{3}{2}T_1,
\end{equation}
where $T_1$ is a given preshock temperature.
In this model, $\beta$ is given by
\begin{equation}
\beta = \frac{T_{\rm vir}}{T_{\rm vir}+ (3/2)T_1}.  \label{eq:betadef}
\end{equation}
If $T_1$ is not negligible compared to $T_{\rm vir}$, $\beta$ becomes
smaller, which means that hot gas expands and the gas density in the
core decreases under the condition that the total gas mass of the cluster
is constant.  As a result, X-ray luminosity becomes smaller, which
results in a steeper $L-T$ relation.  Comparison with observations of
clusters suggests that the preheated temperature, $T_1$, is about 
$0.5-2\ \mbox{keV}$ \citep{FT00}. Here, we adopt $T_1 = 1\ \mbox{keV} $.

Using equation (\ref{eq:Bfit}), we define the normalized central gas 
density as 
\begin{equation}
f_d(\beta) \equiv \frac{\rho_{\rm gas,0}}{\rho_f}
= \frac{4 \pi f_m}{3 I_1(\beta)} \left(\frac{r_f}{r_c} \right)^{3}\:,
\label{eq:gasfrac-p}
\end{equation}
where
\begin{equation}
I_1 (\beta) \equiv 4 \pi \int_{0}^{r_f/r_c} \frac{x^2}{(1+x^2)^{3 \beta/2}} dx.
\end{equation}
From equations (\ref{eq:X-lum}), (\ref{eq:Bfit}) and  (\ref{eq:gasfrac-p}),
the luminosity is written, using $f_d(\beta)$, as
\begin{equation}
L = 3.2 \times 10^{40} \left(\frac{k_B T}{1 \ \mbox{keV}} \right)^{1/2}
\left(\frac{M_f}{10^{15} M_{\odot}} \right) \left[ h^{2}
\Delta_c \frac{\Omega_0}{\Omega(z_f)} \right] (1+z_f)^{3} f_d(\beta)^{2}
\ I_2(\beta) \ \mbox{erg s}^{-1},
\end{equation}
where we assume $r_f/r_c = 8$, and
\begin{equation}
I_2 (\beta) \equiv 4 \pi \int_{0}^{8} \frac{x^2}{(1+x^2)^{3 \beta}} dx.
\end{equation}

Because $\beta$ is a function of $T_{\rm vir}$ once $T_1$ is fixed,  
$T$ $L$ are  
function of $z_f$ and $M_f$ in this preheating model, as is the case of
the scaling model.  Thus, we can calculate the $z_f$ dependences of the
temperature and luminosity functions using the formulation constructed
in {\S} \ref{TL}. The $z_f$ dependence of temperature, 
$T=T(z_f,M = 10^{15} M_{\odot})$, and that of the luminosity, 
$L=L(z_f,M = 10^{15} M_{\odot})$, are shown by thick lines in
Figures \ref{T-z} and \ref{L-z}, respectively. Because of 
preheating, the gas temperature becomes higher and luminosity becomes 
lower compared with the scaling model predictions for this case. 
The temperature and luminosity
functions are shown in Figures \ref{nTp} and \ref{nLp}, respectively.
For several values of $n[\log(T),\log(L);z=0]$, iso-density contours on the 
$L-T$ plane are plotted in Figure \ref{L-Tp}.  
Because of the preheating, the gas
temperature is raised and the gas expands compared to the case without
preheating.  Thus, the central gas density, $f_d$, is
decreased and luminosity is lowered.  When $T_{\rm vir} \sim
T_1$, the effect of preheating is large and luminosity is greatly
decreased. On the other hand, when $T_{\rm vir} \gg T_1$, a cluster is not
much affected by the preheating and $\beta \sim 1$.  Since we assumed
$\beta = 2/3$ in the scaling model, the luminosity in the preheating
model is larger than that in the scaling model in spite of
preheating. Therefore, the slope of $L-T$ relation in the preheating
model is steeper than $L \propto T^{2}$ (Figure \ref{L-Tp}), the number
of clusters with $L > 10^{44} \mbox{erg s}^{-1}$ increases (Figure
\ref{nLp}), and the number of clusters with $L < 10^{44} \mbox{erg
s}^{-1}$ decreases (Figure \ref{nLp}). In the preheating model, there is
also little evolution of luminosity function. This reason is the same
as the scaling model.

In Figures \ref{nTp} and \ref{nLp}, the thick lines are the observed 
temperature and luminosity functions which
are the same as Figure \ref{nT} and Figure \ref{nL}. Both in the SCDM 
and LCDM model, the slope of the luminosity function based on the SSM 
formalism are near to the observed one, 
although it is still slightly steeper in the LCDM model. On the other
hand, the slope of the temperature function is steeper than the observed
one. Better data and better preheating models are needed to resolve 
this mismatch. 
In Figure \ref{L-Tp}, the thick solid lines show the PS prediction 
which becomes as steep as the observed slope because of the preheating 
effect. The thick dotted lines represent the observed $L-T$
relation obtained by \citet{oLT}. Note that observed $L-T$ relation 
has a large dispersion which is comparable to the predicted width. 
Both in the SCDM and LCDM models, 
the predicted $L-T$ relation can match the observed relation rather well.

\section{Conclusions}\label{Conclusion}

We have investigated the effects of formation epoch distribution on
the statistical properties of galaxy clusters in the context of 
hierarchical structure formation scenario. The mass and formation redshift 
of galaxy clusters constitute two independent parameters.  
In this way, using the
formalism of \citet{SSM}, with a few plausible  assumptions,
we have derived temperature and luminosity functions and 
the distribution on the $L-T$ plane. 
First, we investigated a simple scaling model in which gas temperature 
is equal to the virial temperature. The temperature
function is almost the same as that in the PS formalism,
while the luminosity function is shifted to higher luminosity and 
shows no significant evolution independent of the cosmological model 
because earlier formed clusters have denser intracluster gas in the cluster
core. The luminosity-temperature relation becomes a band with a broad width 
instead of a linear line, but its slope becomes a little flatter than that of
the PS formalism, which is inconsistent with the observations.  
Second,
we have examined a simple preheating model in this framework.
Preheating makes the gas distributions of poor clusters flatter than
those of rich clusters, which reduces the X-ray luminosity.  
The resultant $L-T$ relation is steeper than
that in the scaling model and becomes consistent with observations.
Although the temperature and luminosity functions are broadly consistent with 
observations, too, better observations and better preheating models are needed 
for quantitative comparisons.


\clearpage
\figcaption[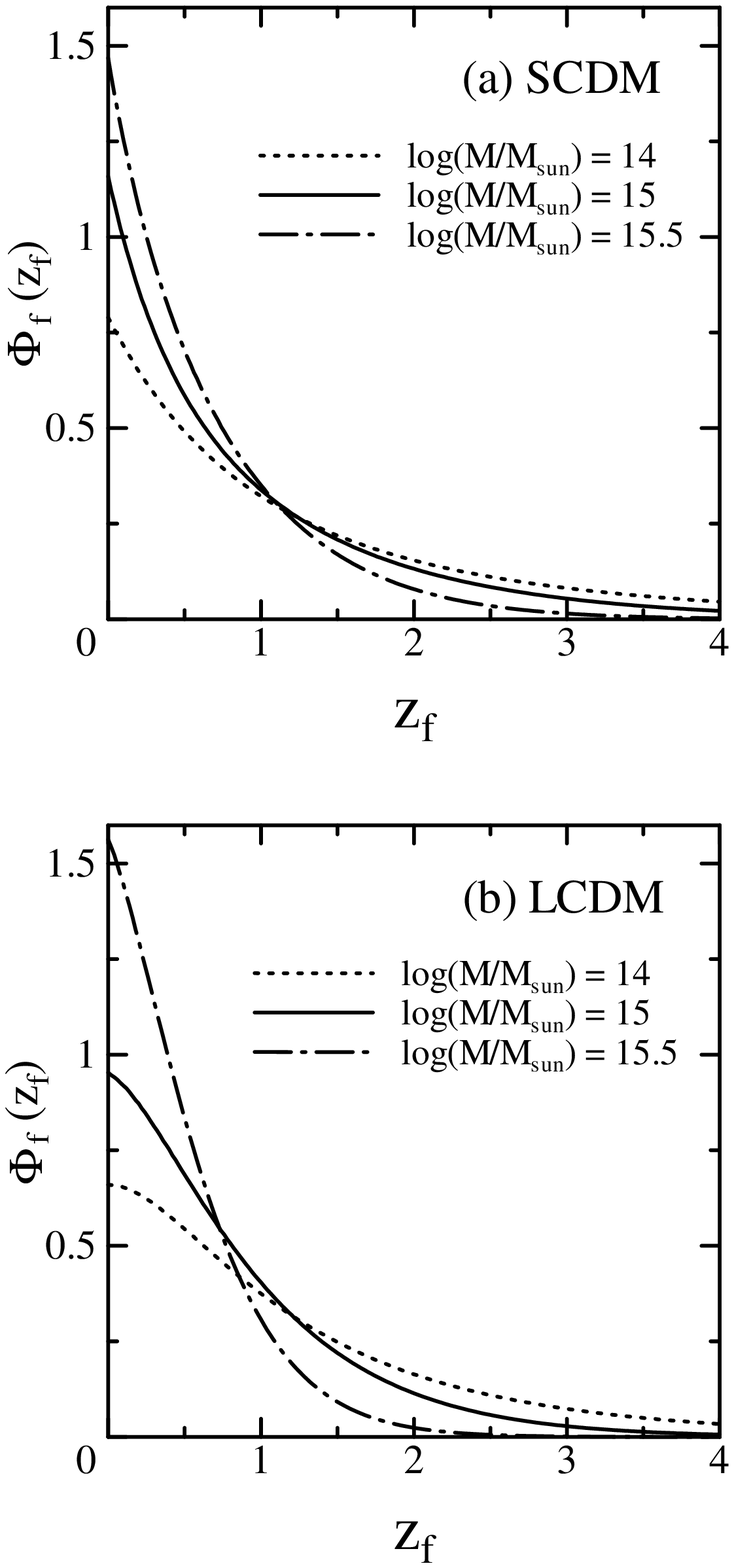]{Formation redshift distribution function, 
$\Phi (z_f;M,z=0)$, for several values of the present halo mass  
in (a) the SCDM model and (b) the LCDM model. \label{dist}} 

\figcaption[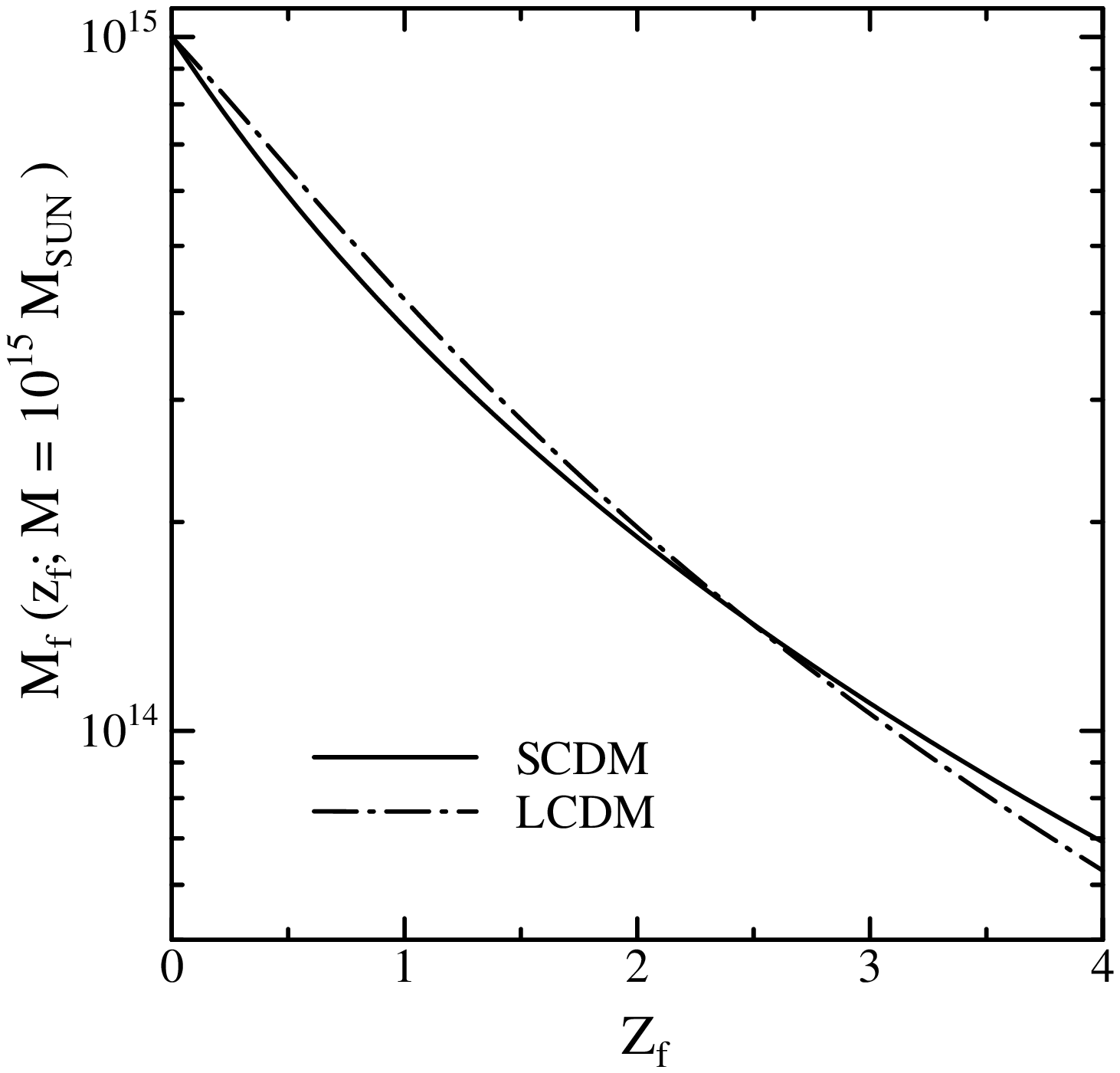]{Halo mass at the formation redshift, 
$M_f(z_f;M=10^{15} M_{\odot})$,
in the SCDM model and the LCDM model. \label{mass}} 

\figcaption[]{Temperature function at $z=0$ and $z=1$ in 
(a) the SCDM model and (b) the LCDM model. 
Symbols indicate the predictions based on the PS formalism, while 
thin lines indicate the predictions based on the SSM formalism. 
Thick line represents the power-law fit to the observed low 
redshift temperature function obtained by \citet{OnT}. 
\label{nT}}

\figcaption[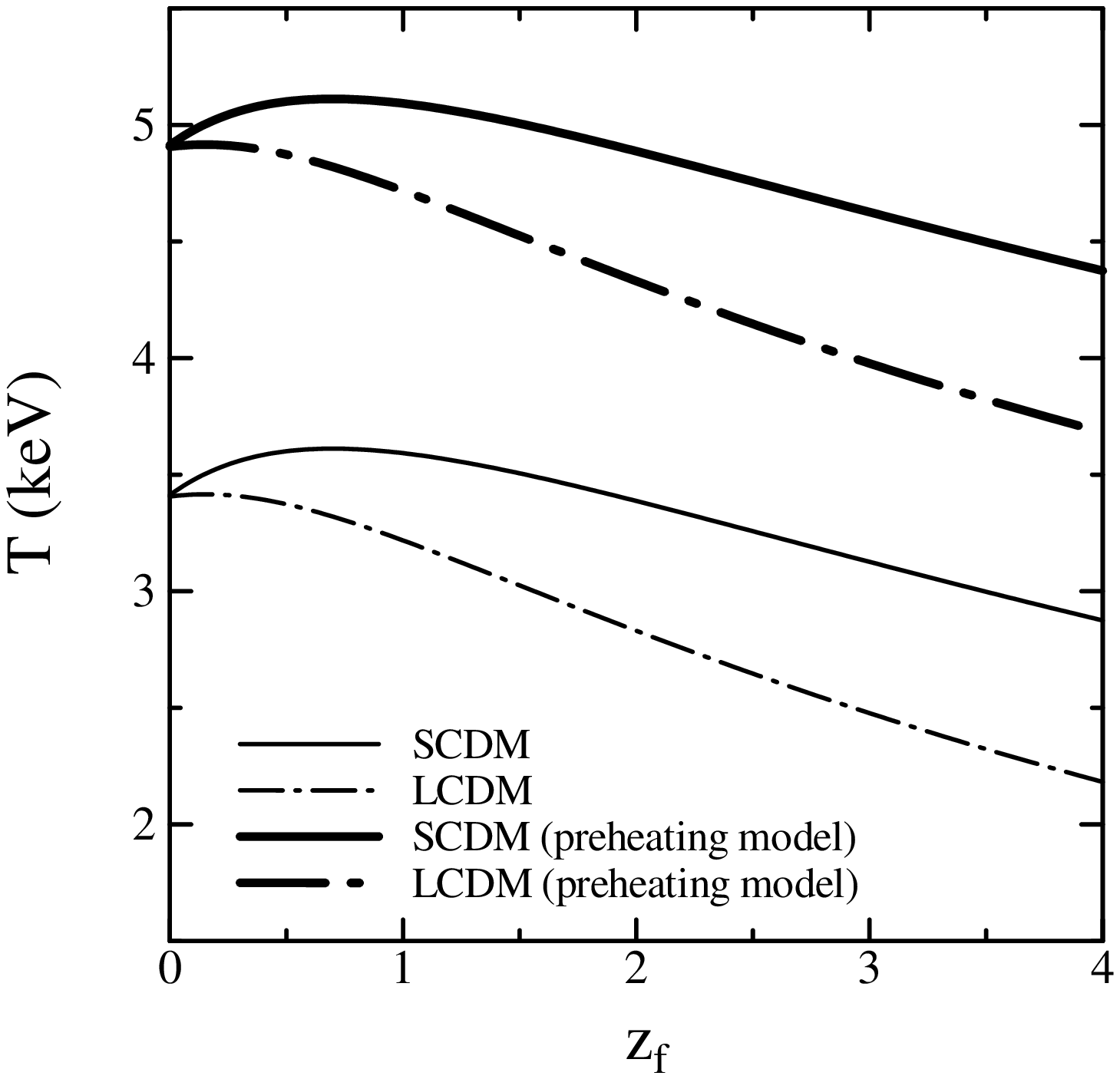]{The temperature-formation redshift relation: 
$T=T(z_f;M=10^{15} M_{\odot})$. Thin lines are the relation in 
the scaling model, while thick lines are the relations in the preheating model. 
\label{T-z}}

\figcaption[]{Luminosity function at $z=0$ and $z=1$ in (a) the SCDM 
and (b) the LCDM. Symbols indicate the predictions based on the PS formalism, 
while thin lines indicate the prediction based on the SSM formalism. 
Thick line represents the best-fitting Schechter function to the 
observed bolometric luminosity function for galaxy clusters within 
$z=0.3$ obtained by \citet{OnL}. \label{nL}}

\figcaption[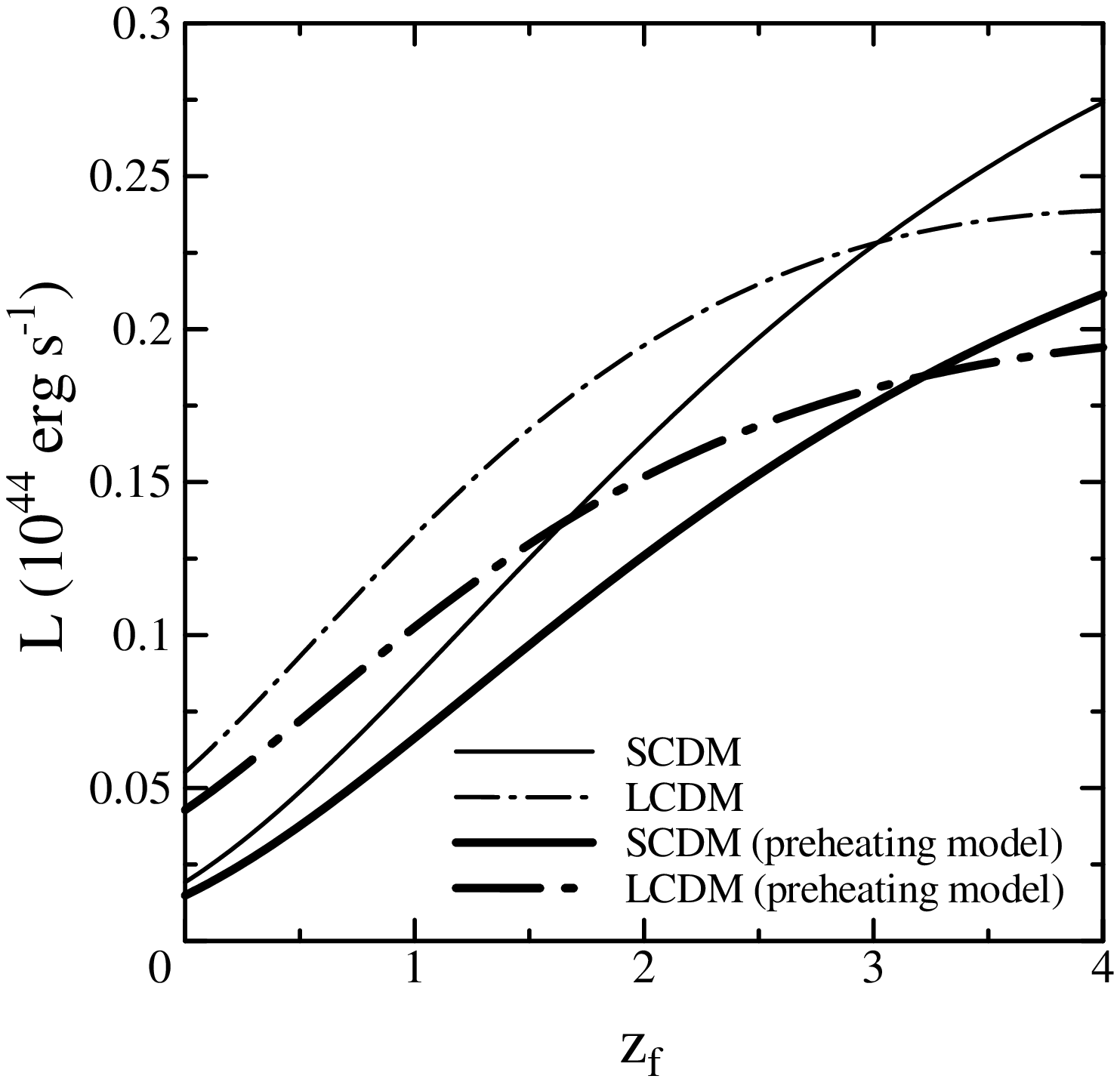]{The luminosity-formation redshift relation: 
$L=L(z_f;M=10^{15} M_{\odot})$.  
Thin lines are the relation in the  scaling model, while thick lines 
are the relations in the preheating model. \label{L-z}}

\figcaption[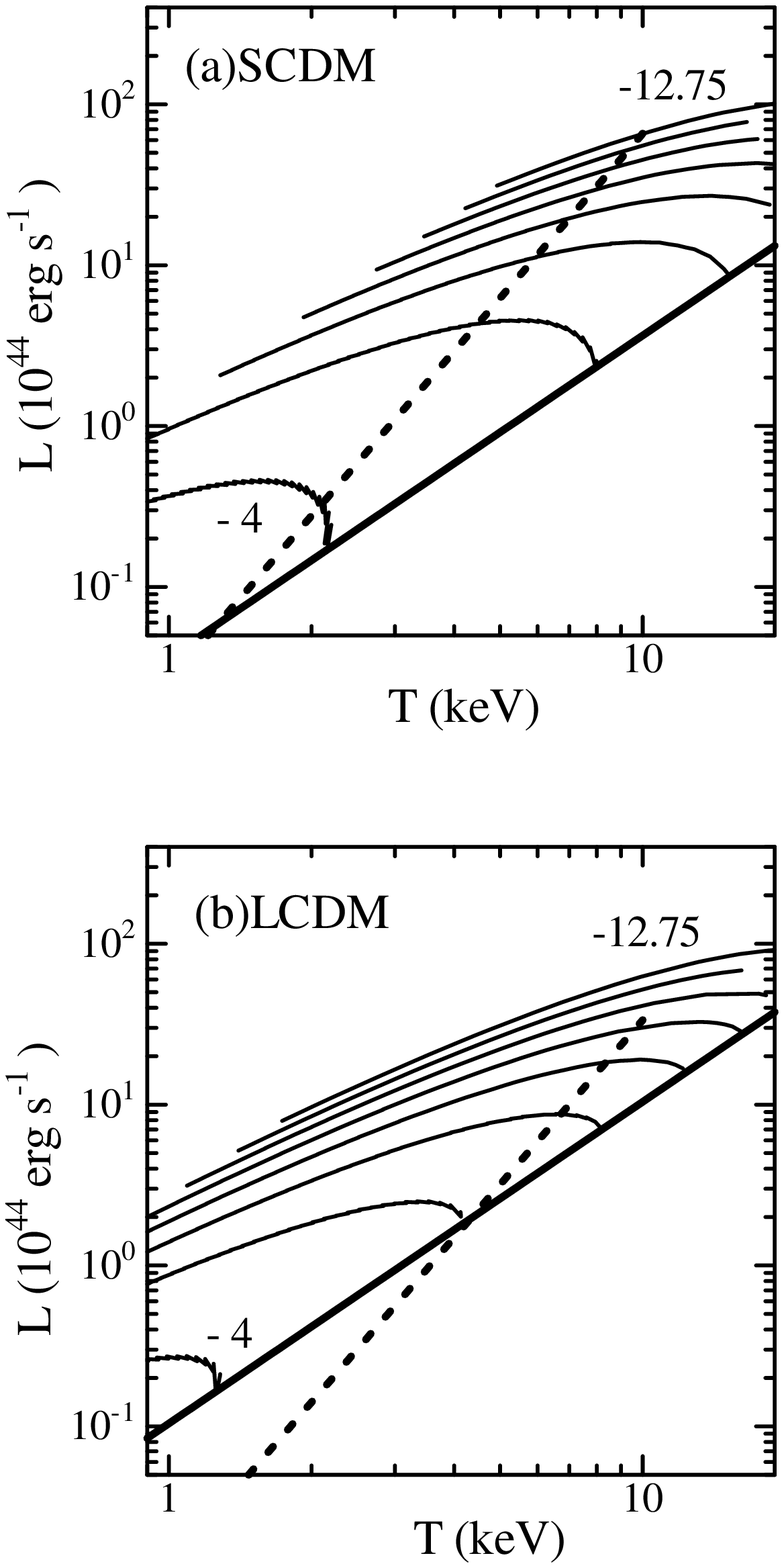]{Iso-density contours on $L-T$ plane in 
(a) the SCDM model and (b) the LCDM model.  
The values of $\log \{n[\log(T), \log(L); z=0]\}$ are separated at equal
logarithmic intervals by 1.25 and range from $-4$ to $-12.75$. Thick
line indicates the predicted $L-T$ relation based on the PS formalism.
This corresponds to the case $z_f = 0$ and follows $L \propto
T^2$. Dotted line represents the power-law fit to the observed $L-T$
relation obtained by obtained by \citet{oLT}. \label{L-T}}

\clearpage

\figcaption[]{Temperature function in the preheating model at 
$z=0$ and $z=1$ in (a) the SCDM model and (b) the LCDM model. 
Thick line represents the power-law fit to the observed low redshift 
temperature function obtained by \citet{OnT}. \label{nTp}}

\figcaption[]{Luminosity function in the preheating model at $z=0$ 
and $z=1$ in (a) the SCDM model and (b) the LCDM model. 
Thick line represents the best-fitting Schechter function to the 
observed bolometric luminosity function within $z=0.3$ obtained by 
\citet{OnL}. \label{nLp}}

\figcaption[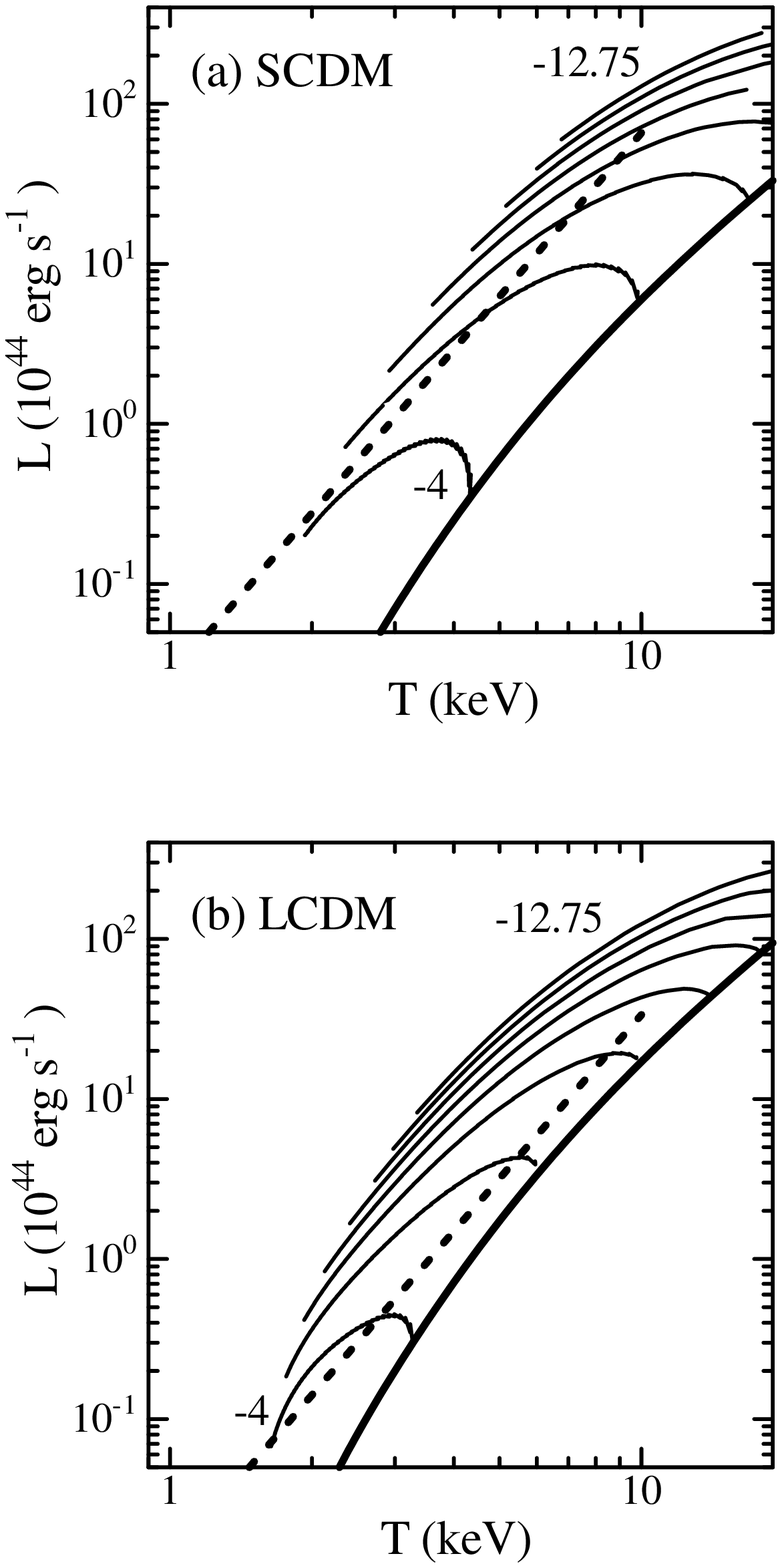]{Iso-density contours on the $L-T$ plane 
in (a) the SCDM model and (b) the LCDM model.  
The values of $\log \{n[\log(T), \log(L); z=0]\}$ are separated at equal
logarithmic intervals by 1.25 and range from $-4$ to $-12.75$.
Thick line indicates the predicted $L-T$
relation based on the PS formalism.  This corresponds to the case of
$z_f = 0$. 
Dotted line represents the power-law fit to the observed $L-T$
relation obtained by obtained by \citet{oLT}. \label{L-Tp}}


\clearpage

\begin{table}
\caption{Cosmological parameters of the models} \begin{center}
\begin{tabular}{lccccc} \hline\hline
Model & $\Omega_0$ & $\lambda_0$ & $h$ & $\sigma_8$  \\
\hline SCDM & 1.0 & 0.0 & 0.5 & 1.0  \\ LCDM & 0.3 & 0.7 & 0.7 & 1.0  \\
\hline
\end{tabular}
\end{center}
\label{parameter}
\end{table}%

\end{document}